\pdfoutput=1
\documentclass[a4paper,11pt]{article}

\usepackage{amsmath}
\usepackage{amssymb}
\usepackage{graphicx}
\graphicspath{{figures/}} 
\usepackage{xfrac}

\usepackage{braket}

\usepackage[utf8]{inputenc}

\usepackage[square,numbers,comma,sort&compress]{natbib}
\usepackage[colorlinks=true,citecolor=blue,linkcolor=purple,urlcolor=purple]{hyperref}
\usepackage{geometry}
\usepackage[capitalise]{cleveref}
\usepackage[usenames,dvipsnames,svgnames,table]{xcolor}
\usepackage{booktabs}
\usepackage{slashed}

\def\varabstract{ }
\def\varkeywords{ }
\def\vararxivnumber{ }
\def\vartitle{ }
\def\varsubtitle{ }
\renewcommand{\title}[1]{\gdef\vartitle{#1}}

\renewcommand{\abstract}[1]{\gdef\varabstract{#1}}
\newcommand{\keywords}[1]{\gdef\varkeywords{#1}}

\newtoks\authtoks
\renewcommand{\author}[2][]{%
	\authtoks=\expandafter{\the\authtoks#2$^{#1}$\ }%
}
\newtoks\affiltoks
\newcommand{\affiliation}[2][]{%
    \affiltoks=\expandafter{\the\affiltoks{\item[$^{#1}$]#2}}%
}
\newtoks\emailtoks\newcounter{emailcounter}%
\newcommand{\emailAdd}[1]{%
\ifnum\theemailcounter>0\emailtoks=\expandafter{\the\emailtoks, \typeemail{#1}}%
\else\emailtoks=\expandafter{\typeemail{#1}}%
\fi
\stepcounter{emailcounter}}
\newcommand{\typeemail}[1]{\href{mailto:#1}{\tt #1}}

\renewcommand\maketitle{
	\newgeometry{margin=2cm}
	\pagestyle{empty}\setcounter{page}{0}
	{\huge\flushleft\sffamily\bfseries\vartitle\\\Large\varsubtitle\par}
\vskip6ex
{\large\bfseries\raggedright\sffamily\the\authtoks\par}
\vskip2ex
\begin{list}{}{%
\setlength{\leftmargin}{0.28cm}%
\setlength{\labelsep}{0pt}%
\setlength{\itemsep}{-3pt}%
\setlength{\topsep}{-\parskip}}
\itshape\small%
\the\affiltoks
\end{list}
\vskip2ex
\noindent\hspace{0.28cm}\begin{minipage}[l]{.9\textwidth}
\begin{flushleft}
\textit{E-mail:} \the\emailtoks
\end{flushleft}
\end{minipage}
\vskip5ex
\noindent{\renewcommand\baselinestretch{.9}\textsc{Abstract:}}\ \varabstract
\vskip5ex
\if!\varkeywords!\else\noindent{\textsc{Keywords:}}\ \varkeywords \vskip2ex\fi
\if!\vararxivnumber!\else\noindent{\textsc{ArXiv ePrint:}} \href{http://arxiv.org/abs/\vararxivnumber}{\vararxivnumber}\vskip2ex\fi

\newpage
\restoregeometry
\pagestyle{plain}

\setcounter{footnote}{0}
}

\usepackage[square,numbers,comma,sort&compress]{natbib}

\definecolor{MS}{rgb}{0,0,1}

\usepackage{ifthen}

	\newcommand{\barlimc}[7]{
  \pgfmathparse{\mypos+0.3}
  \edef\mypos{\pgfmathresult}
		\node[left,scale=0.6] at (0,\mypos) {#1};
		\pgfmathparse{#3 > 5 ? 1 : 0}
		\ifthenelse{\pgfmathresult=1}{
			\fill[#2] ($(0,\mypos)+(0,-0.1)$) rectangle +(5,0.2);
			\fill[white] ($(0,\mypos)+(3.5,-0.1)$) rectangle +(0.3,0.2);
			\draw[decoration={zigzag},decorate,#2,very thick] (3.4,\mypos) to +(0.5,0);
			\node[left,scale=0.6] at (5,\mypos) {#3};
			}{
			\fill[#2] ($(0,\mypos)+(0,-0.1)$) rectangle +(#3,0.2);
			\node[left,scale=0.6] at (#3,\mypos) {#3};
		}
		\fill[#4] ($(0,\mypos)+(0,-0.1)$) rectangle +(#5,0.2);
		\node[left,scale=0.6] at (#5,\mypos) {#5};
		\fill[#6] ($(0,\mypos)+(0,-0.1)$) rectangle +(#7,0.2);
		\pgfmathparse{#7 <0.3 ? 1 : 0}
		\ifthenelse{\pgfmathresult=1}{
			\node[right,scale=0.6] at (0,\mypos) {#7};
		}{
		\node[left,scale=0.6] at (#7,\mypos) {#7};
	}
}

\title{$R^\nu_{K^{(*)}}$ and non-standard neutrino interactions}
\author[1,2]{Xiao Gang He}\emailAdd{hexg@phys.ntu.edu.tw}
\author[3]{and German Valencia}\emailAdd{German.Valencia@monash.edu}

\affiliation[1]{
Tsung-Dao Lee Institute, and School of Physics and Astronomy, Shanghai Jiao Tong University, Shanghai 200240}

\affiliation[2]{Department of Physics, National Taiwan University, Taipei 10617}

\affiliation[3]{School of Physics and Astronomy, Monash University, Wellington Road, Clayton, VIC-3800, Australia}

\abstract{We discuss the modes $B\to K^{(*)}\nu\bar\nu$ in the context of non-standard neutrino interactions that add incoherently to the SM rates. We consider two scenarios: an additional light neutrino; and neutrino lepton flavour violation. We find that an additional light neutrino that interacts with SM fields via a non-universal $Z^\prime$ can increase $R^\nu_{K^{(*)}}$ by up to a factor of two without conflicting with $B_s-\bar B_s$ mixing. This model then predicts rates for $B_s\to \tau^+\tau^-$ up to six times larger than the SM. In the context of neutrino lepton flavour violation mediated by leptoquarks we find that the current experimental upper bounds on $R^\nu_{K^{(*)}}$ are already more constraining than direct bounds from $B_s\to \tau \ell$ and $B\to  K^{(*)}\tau \ell$ modes for $\ell=e,\mu$.  

}

\keywords{}

\begin{document}

\maketitle


\newpage

\section{Introduction}\label{s:intro}

Rare $B$ decays play an important role in understanding the dynamics of the standard model (SM) as well as being a fertile ground for the search for new physics. The $B\to K^{(*)} \nu \bar \nu$ decays are amongst the cleanest modes to search for new physics due to their well controlled theoretical uncertainty. Experiments at Belle and Babar have already published upper limits on these modes at 2-3 times the SM rate and further improvement is expected from Belle-II, which can reach a sensitivity on the branching ratios of about 10\% with 50 ab$^{-1}$ \cite{Belle-II:2018jsg}. Interesting constraints for certain BSM physics can be  obtained already with current bounds. 

We consider two types of models that add incoherently to the SM rates. First we entertain the possibility of a fourth light neutrino that couples to SM fields through a non-universal $Z^\prime$. We find that existing constraints on the model allow enhancements of the $B\to K^{(*)} \nu \bar \nu$ rates by up to factors of two and that these are correlated with the $B_s\to \tau^+\tau^-$ mode which could reach a rate up to six times larger than in the SM.

We then consider possible contributions from neutrino flavour violating final states in the context of scalar and vector leptoquarks. These contributions are correlated to the charged lepton flavour violating (CLFV) modes $B_s \to \ell\ell^\prime$ and $B\to K^{(*)}\ell \ell^\prime$ and we find that the current limits on $B\to K^{(*)} \nu \bar \nu$ are more restrictive for modes with tau-leptons. The leptoquark scenario also correlates  $B\to K^{(*)} \nu \bar \nu$ to the neutral and charged B anomalies, as has been extensively discussed in the literature, and we comment on this.

Within the SM the effective Hamiltonian responsible for the $B\to K^{(*)} \nu \bar \nu$ transitions originates at lowest order from box and penguin diagrams and is usually written as \cite{Buchalla:1995vs}
\begin{align}
{\cal H}_{SM}&=-\frac{4G_F}{\sqrt{2}}V_{tb}V^\star_{ts}C_L^{SM}\sum_i{\cal O}^{ii}_L+{\rm~h.c.}  \nonumber \\
{\cal O}^{ii}_L &= \frac{e^2}{16\pi^2}(\bar s\gamma_\mu P_Lb)(\bar\nu_i\gamma^\mu(1-\gamma_5)\nu_i), 
\label{smeh}
\end{align}
with an accurately known  Wilson coefficient that is independent of the neutrino flavour and that including NLO QCD corrections \cite{Buchalla:1998ba} and two-loop electroweak corrections \cite{Brod:2010hi} is given by
\begin{align}
C_L^{SM}=-\frac{X(x_t)}{s^2_W},\quad X(x_t)=1.469\pm0.017.
\label{smwc}
\end{align}
Typical SM predictions obtained with {\tt flavio} are\footnote{These numbers agree within errors with published numbers as in \cite{Blake:2016olu,Belle-II:2018jsg}. Neglecting isospin breaking, the neutral and charged modes have the same rates so we choose to present the two modes with the strongest experimental limits.}
\begin{align}
\mathcal{B}(B^+\to K^+\nu\bar\nu)_{SM}=(4.4\pm 0.7)\times 10^{-6},\quad \mathcal{B}(B^0\to K^{*0}\nu\bar\nu)_{SM} = (9.5\pm 1.0)\times 10^{-6}.
\label{smrates}
\end{align}

We list the best current experimental constraints on these modes in \cref{tab:expbounds}.
\begin{table}[h]
	\centering
	\begin{tabular}{|l|l|l|}\toprule
		\textbf{Mode} & \textbf{90\% c.l upper limit} & \textbf{Reference}  \\ \midrule
		$\mathcal{B}(B^+\to K^+\nu\bar\nu)$ & $1.6\times 10^{-5}$& Babar \cite{Lees:2013kla} \\
		$\mathcal{B}(B^+\to K^{* +}\nu\bar\nu)$ & $4.0\times 10^{-5}$ & Belle \cite{Lutz:2013ftz} \\
		$\mathcal{B}(B^0\to K^0\nu\bar\nu)$ & $2.6\times 10^{-5}$   & Belle \cite{Grygier:2017tzo} \\
		$\mathcal{B}(B^0\to K^{* 0}\nu\bar\nu)$ & $1.8\times 10^{-5}$   & Belle \cite{Grygier:2017tzo} \\
		 \bottomrule
	\end{tabular}
		\caption{Current experimental upper bounds on the modes considered.}
	\label{tab:expbounds}
\end{table}
Belle-II is expected to improve these limits, and has produced a preliminary result  $\mathcal{B}(B^+\to K^+\nu\bar\nu)\leq 4.1\times 10^{-5}$ at the 90\% confidence level~\cite{Dattola:2021cmw}. They have averaged this result  with the previous ones to arrive at $\mathcal{B}(B^+\to K^+\nu\bar\nu)=(1.1\pm 0.4)\times 10^{-5}$~\cite{Dattola:2021cmw}. For the $K^*$ channel, Belle has also combined the charged and neutral modes to obtain the limit $\mathcal{B}(B\to K^*\nu\bar\nu)\leq 2.7\times 10^{-5}$ \cite{Grygier:2017tzo} but here we will use the limit in \cref{tab:expbounds}. These results are usually presented as ratios, for which we obtain
 \begin{align}
R^\nu_K = \frac{\mathcal{B}(B^+ \to K^+ \nu \bar \nu)}{\mathcal{B}(B^+ \to K^+ \nu \bar \nu)_{SM}} = 2.5\pm 1.0, \quad
R^\nu_{K^{*}} = \frac{\mathcal{B}(B \to K^{*0} \nu \bar \nu)}{\mathcal{B}(B\to K^{*0} \nu \bar \nu)_{SM}} \leq  1.9.
\label{Rratios}
\end{align}
The second number is simply the ratio of the limit in \cref{tab:expbounds} and the central value in \cref{smrates} and somewhat lower than what is used in \cite{Browder:2021hbl}.

\section{Effective Hamiltonian at the $b$ scale}

We can parameterise any new physics relevant for these decays through an effective Hamiltonian at the $b$ mass scale. The effective theory originates in extensions of the SM containing new particles at or above the electroweak scale that have been integrated out. In general, this results in additional contributions to $C_L$ in \cref{smwc} as well as in new operators. Because our discussion is tied to two types of models, we only need to consider the following
\begin{align}
{\cal H}_{NP} = -\frac{4G_F}{\sqrt{2}}V_{tb}V^\star_{ts}\frac{e^2}{16\pi^2}
&\sum_{ij}\left(C_{L}^{ij}{\cal O}_L^{ij}+C_{R}^{ij}{\cal O}_R^{ij}+
C_{L}^{\prime~ij}{\cal O}_L^{\prime~ij}+C_{R}^{\prime~ij}{\cal O}_R^{\prime~ij}\right. \nonumber \\
&+\left. C_9^{ij} {\cal O}_9^{ij}+C_{10}^{ij} {\cal O}_{10}^{ij}+C_{9^\prime}^{ ij} {\cal O}_{9^\prime}^{ ij}+C_{10^\prime}^{ ij} {\cal O}_{10^\prime}^{ ij}\right)+{\rm ~h.c.}
\label{effHb}
\end{align}
where the operators are
\begin{align}
{\cal O}_L^{ij}&= (\bar s_L\gamma_\mu b_L)(\bar\nu_i\gamma^\mu(1-\gamma_5)\nu_j)&
{\cal O}_R^{ij}&=(\bar s_R\gamma_\mu b_R)(\bar\nu_i\gamma^\mu(1-\gamma_5)\nu_j) \nonumber \\
{\cal O}_L^{\prime~ij} &=(\bar s_L \gamma_\mu b_L )(\bar\nu_i\gamma^\mu(1+\gamma_5)\nu_j)&
{\cal O}_R^{\prime~ij} &=(\bar s_R\gamma_\mu b_R)(\bar\nu_i\gamma^\mu(1+\gamma_5)\nu_j)\nonumber \\
{\cal O}_{9}^{ij} &=(\bar s_L \gamma_\mu b_L)(\bar\ell_i \gamma^\mu \ell_j)&
{\cal O}_{10}^{ij} &=(\bar s_L \gamma_\mu b_L)(\bar\ell_i \gamma^\mu\gamma_5 \ell_j)\nonumber \\
{\cal O}_{9^\prime}^{ij} &=(\bar s_R \gamma_\mu b_R)(\bar\ell_i \gamma^\mu \ell_j) &
{\cal O}_{10^\prime}^{ij} &=(\bar s_R \gamma_\mu b_R)(\bar\ell_i \gamma^\mu\gamma_5 \ell_j)
\label{eq:ops}
\end{align}
The Wilson coefficients are defined so that they only contain NP contributions and the SM is counted separately through \cref{smwc}. The list in \cref{eq:ops} includes ${\cal O}^{(\prime)}_L$, ${\cal O}^{(\prime)}_R$, which contribute to  $B^{(*)}\to K^{(*)} \nu \bar \nu$. It excludes operators with scalar and tensor neutrino bi-linears that have been considered in \cite{Browder:2021hbl} because they do not appear in the models we discuss. The operators with charged leptons appear in the models we discuss with coefficients that are related to $C_{L,R}^{(\prime)ij}$. 
The flavour diagonal operators ${\cal O}^{(\prime)\mu\mu}_{9, 10}$  affect $b\to s \mu \bar \mu$ decays including the  $B\to K ^{(*)}\mu \bar \mu$ anomalies and have been studied extensively in that context.

We can classify the contributions to $B\to K^{(*)}\nu\bar\nu$ from these operators into two types: those that interfere with the SM, ${\cal O}_{L,R}^{ii}$; and those that do not, ${\cal O}_{L,R}^{i\neq j}$  and ${\cal O}^{\prime~ij}_{L,R}$. 
In $B\to K\nu\bar\nu$ only the vector current enters the hadronic matrix element so that the contributions to the rate from ${\cal O}_L^{ij}$ and ${\cal O}_R^{ij}$ are the same. Similarly for those from ${\cal O}_L^{\prime~ij}$ and ${\cal O}_R^{\prime~ij}$. At the same time, the different neutrino chirality eliminates interference between the primed and un-primed-operators for massless neutrinos. The only operators that interfere with the SM are thus the diagonal ones (in neutrino flavour) ${\cal O}_L^{ii}$ and ${\cal O}_R^{ii}$. The rates can be evaluated numerically using $\tt flavio$ \cite{Straub:2018kue}, and the central value (uncertainty will be shown in the figures) is given approximately by 
\begin{align}
\mathcal{B}(B^+\to K^+\nu\bar\nu) \times 10^6 &\approx 4.39-0.457~{\rm Re}\sum_i\left( C^{ii}_{L}+C^{ii}_R\right)
\nonumber \\
&+0.0357 ~\sum_{ij}\left( \left|C^{ij}_L+C^{ij}_R\right|^2 +\left|C_{L}^{\prime~ij}+C_{R}^{\prime~ij}\right|^2 \right)
\label{scalarbr}
\end{align}

In $B\to K^*\nu\bar\nu$ both the vector and axial-vector currents enter the hadronic matrix element resulting in different contributions for ${\cal O}_L^{ij}$ and ${\cal O}_R^{ij}$ as well as for ${\cal O}_L^{\prime~ij}$ and ${\cal O}_R^{\prime~ij}$. Numerically, the rate is approximately given by
\begin{align}
\mathcal{B}(B^{0}\to K^{\star 0}\nu\bar\nu) \times 10^6 &\approx  9.53+{\rm Re}\sum_i\left( -0.993~C^{ii}_{L}+0.661~C^{ii}_R\right) \nonumber\\
&+  \sum_{ij}\left(0.0775\left({C^{ij}_L}^{2}+{C^{ij}_R}^2 +{C_{L}^{\prime~ij}}^2+{C_{R}^{\prime~ij}}^2\right)-0.103\left(C^{ij}_LC^{ij}_R+C_{L}^{\prime~ij}C_{R}^{\prime~ij}\right)\right)
\label{vectorbr}
\end{align}

\begin{figure}[!htb]
\begin{center}
\centering{\includegraphics[width=6cm]{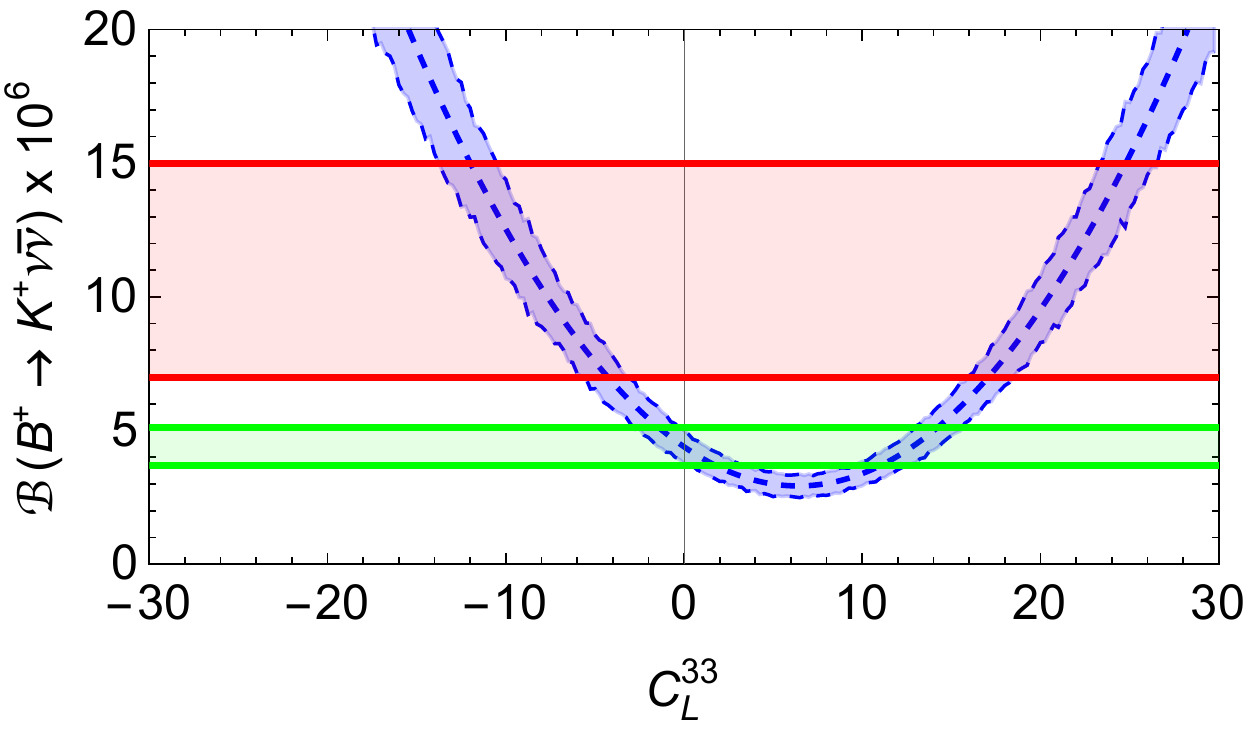}\includegraphics[width=6cm]{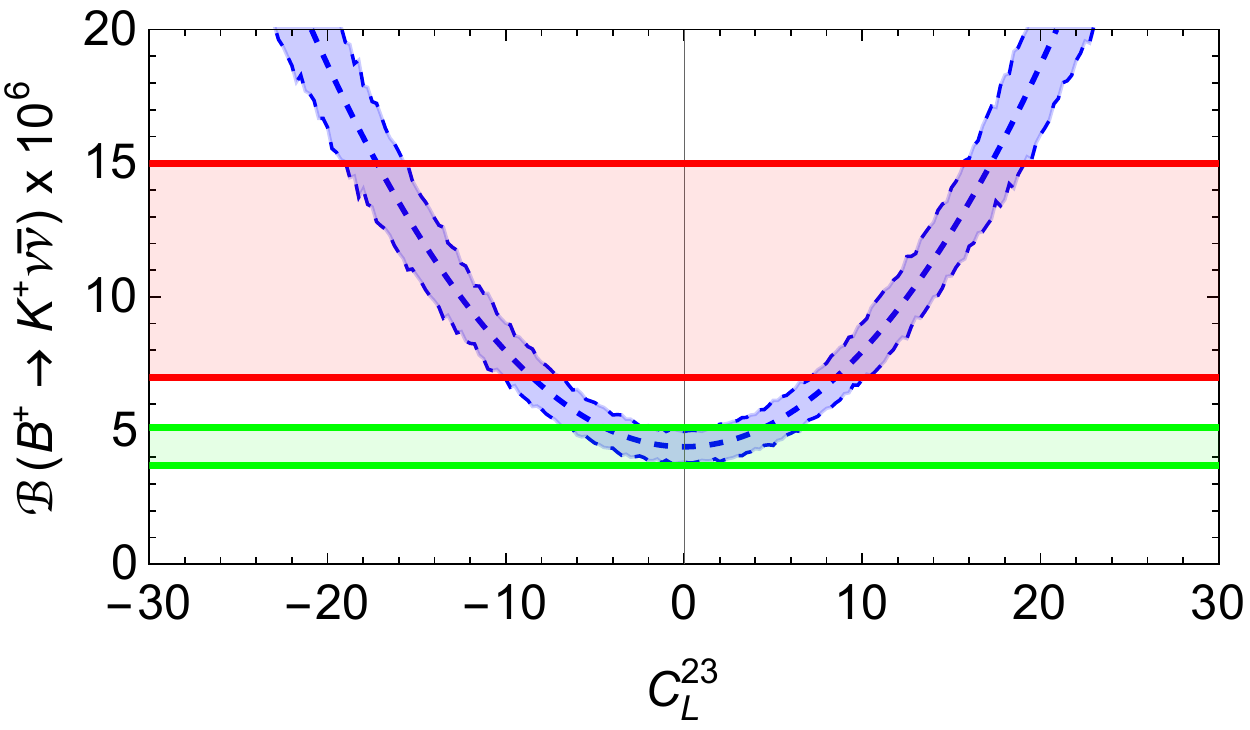}}
\centering{\includegraphics[width=5cm]{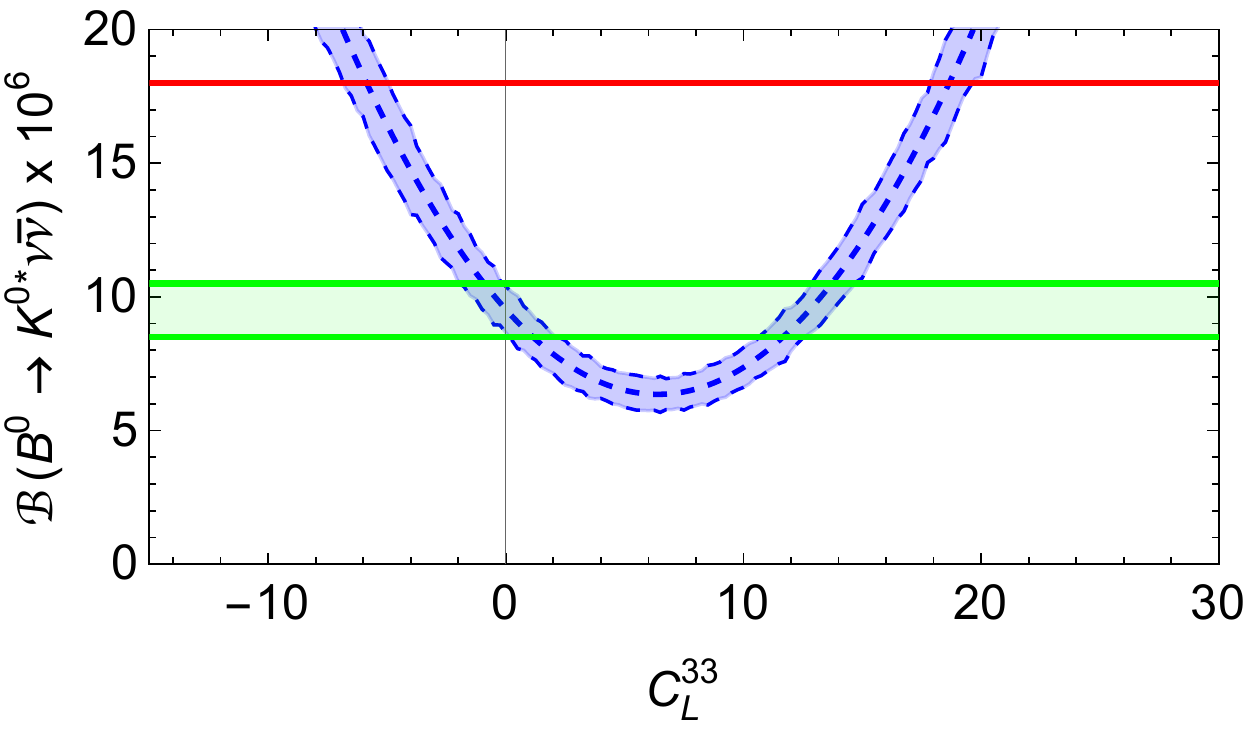}\includegraphics[width=5cm]{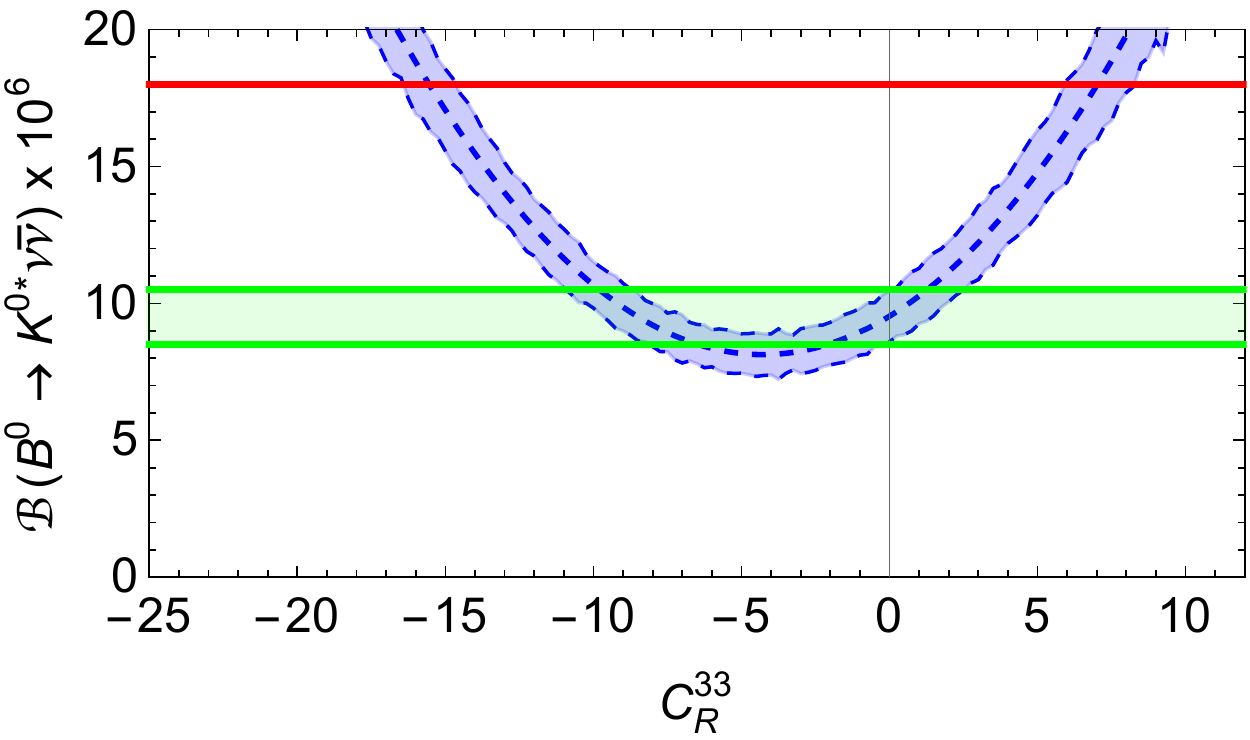}\includegraphics[width=5cm]{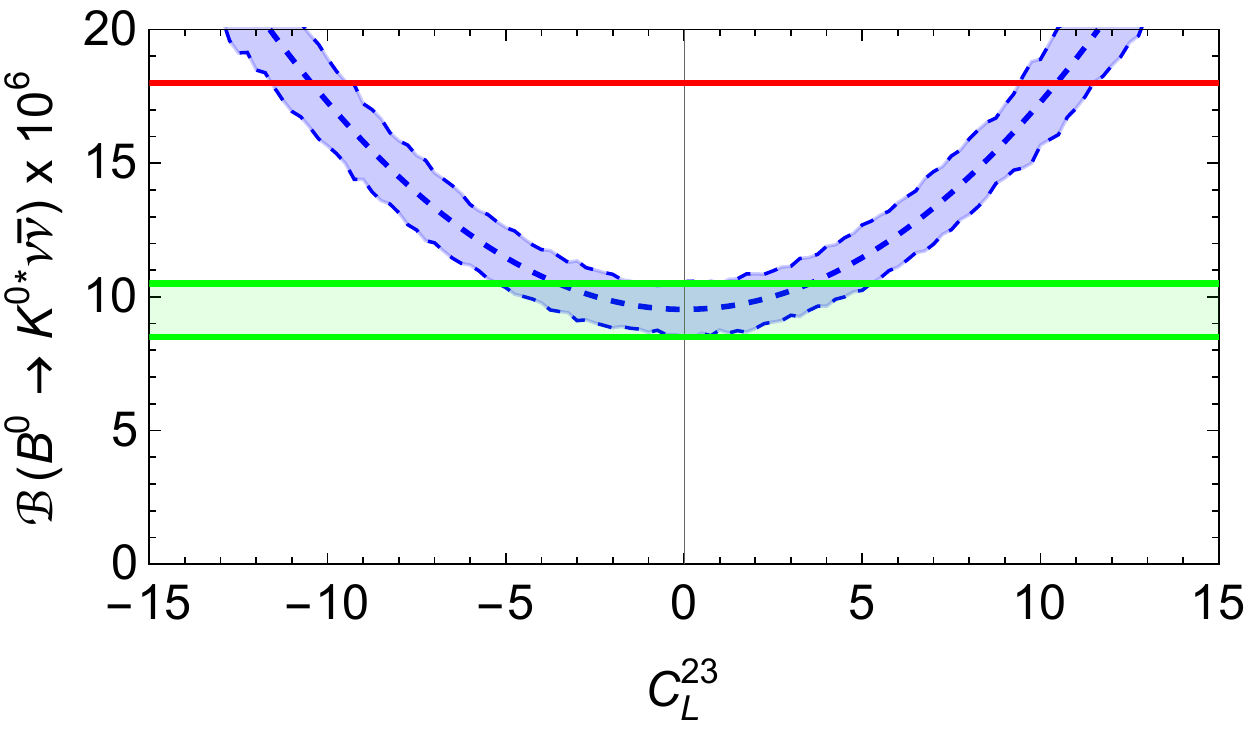}}
\end{center}
\caption{Top row: $\mathcal{B}(B^+\to K^+\nu\bar\nu)$ as a function of $C_{L}^{33}$ (left panel) and $C_{L}^{23}$ (right panel). Bottom row:  $\mathcal{B}(B^{0}\to K^{\star 0}\nu\bar\nu)$ as a function of $C_{L}^{33}$ (left panel),  $C_{R}^{33}$ (centre panel) and $C_{L}^{23}$ (right panel).}
\label{f:bpkp}
\end{figure}

The parametric uncertainty in these predictions, as estimated by {\tt flavio} is illustrated in \cref{f:bpkp}\footnote{The uncertainty in these predictions is around 15\% and is mostly due to the form factors for the $B\to K$ and $B\to K^*$ hadronic transitions, which are responsible for about 10\%, while the value of $V_{cb}$ contributes an additional  5\%.}.  We show $\mathcal{B}(B^+\to K^+\nu\bar\nu)$ as a function of $C_{L}^{33}$ (the figure is identical for $C_{L,R}^{ii}$) and as a function of $C_{L}^{23}$ (the figure is identical for $C_{L}^{i\neq j}$ and $C_{L,R}^{\prime~ij}$). The red band marks the experimental combination $\mathcal{B}(B^+\to K^+\nu\bar\nu)=(1.1\pm 0.4)\times 10^{-5}$~\cite{Dattola:2021cmw} and the green band the SM in \cref{smrates} at 1$\sigma$. For $\mathcal{B}(B^{0}\to K^{\star 0}\nu\bar\nu)$ we show the dependence on $C_{L}^{33}$ (same for all $C_{L}^{ii}$), $C_{R}^{33}$ (same for all $C_{R}^{ii}$) and $C_{L}^{23}$ (same for all $C_{L}^{i\neq j}$ and $C_{L,R}^{\prime~ij}$). In this case the red line shows the 90\% c.l. experimental upper bound from \cref{tab:expbounds} and the green band the SM at 1$\sigma$.

For example, new physics contributions allowed by the value of $\mathcal{B}(B^+\to K^+\nu\bar\nu)$ in \cref{scalarbr} at the 1$\sigma$ level and taking only one non-zero parameter at a time are shown in \cref{tab:milim} and can also be read off \cref{f:bpkp}. 
\begin{table}[h]
	\centering
	\begin{tabular}{|c|c|c|c|}\toprule
		\textbf{Mode} & $C^{ii}_L$ & $C^{ii}_R$ & $|C^{i\neq j}_{L,R}|$ or $|C^{\prime~ij}_{L,R}|$  \\ \midrule
$1.5 \leq R^\nu_K \leq 3.5$ & $[-12,-4]$ or $[17,25]$& $[-12,-4]$ or $[17,25]$ & $[8.5,17]$ \\
$R^\nu_{K^*} \leq 2.2$ & $[-6,18.5]$& $[-15.5,7]$ & $[0,10.5]$  \\
		 \bottomrule
	\end{tabular}
		\caption{Limits for the coefficients $C_{L,R}^{ij(\prime)}$ taken one at a time implied by \cref{Rratios}.}
	\label{tab:milim}
\end{table}

\section{An additional light neutrino}

The first type of new physics we consider that can increase the SM value of $B\to K^{(*)}\nu\bar\nu$ consists of new light neutrinos. In fact, modes with neutrino pairs in the final state count the number of light neutrinos in the SM due to lepton universality. The existence of new light neutrinos is severely constrained by measurements of the invisible $Z$ width and by cosmological considerations. Assuming lepton universality, the former implies that $N_\nu =2.9840\pm0.0082$ \cite{ALEPH:2005ab}. Cosmological constraints depend on other parameters and, for example,  $\Delta N_{eff} < 0.77$ for a Hubble constant $H_0=71.3^{+1.9}_{-2.2}$~km/s/Mpc \cite{Bernal:2016gxb}. 

It is possible to avoid these limits with a light sterile neutrino that interacts with the SM through a $Z^\prime$. The contribution of this neutrino to the $Z$ width neutrino count is proportional to the square of the $Z-Z^\prime$ mixing parameter and can thus be negligibly small.  In addition, if the $Z^\prime$ is non-universal and couples predominantly to the third generation SM fermions, the new neutrino reaches thermal equilibrium with SM particles at a temperature near the $\tau$-lepton mass.  However, at the time of big-bang nucleosynthesis the temperature is about 1~MeV and this difference results in a suppression of the contribution of this neutrino to $\Delta N_{eff}$ to a safe level, $\Delta N_{eff}<0.1$ \cite{Dolgov:2002wy}. 

\begin{figure}[!htb]
\begin{center}
\centering{\includegraphics[width=7cm]{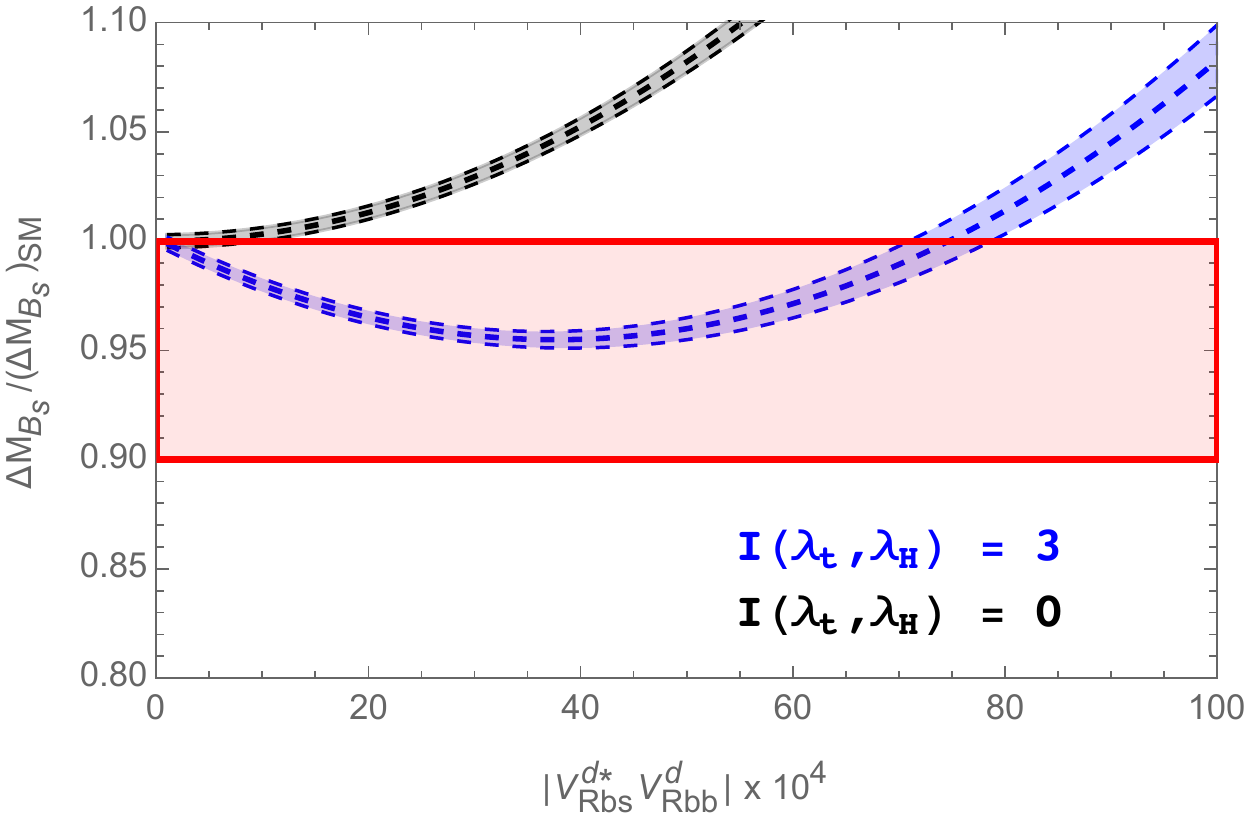}\includegraphics[width=8cm]{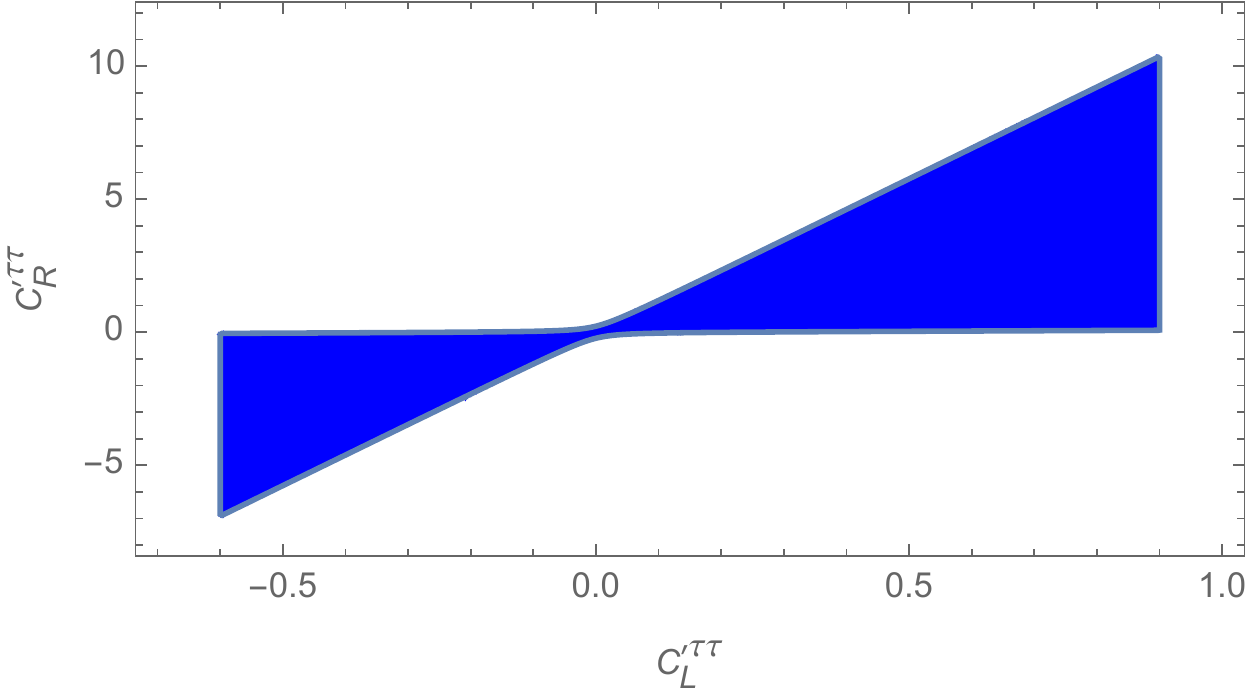}}
\end{center}
\caption{Left panel: $\Delta M_{B_s}/(\Delta M_{B_s})_{SM}$ as a function of $|V^d_{Rbs} V^{d*}_{Rbb} |$ for  $I(\lambda_t,\lambda_H)=0,3$ showing parametric uncertainty. Right panel: values taken by $C_{L,R}^{\prime\tau\tau}$ with parameters that satisfy $0.9\leq \Delta M_{B_s}/(\Delta M_{B_s})_{SM}\leq 1.0$ assuming that $V^d_{Rbs} V^{d*}_{Rbb}$ is real. }
\label{f:bsmix}
\end{figure}

We have previously constructed a detailed example of a model with these properties \cite{He:2002ha,He:2017bft} so we do not repeat the details here. The $Z^\prime$ is responsible for two new operators that contribute to $B \to K^{(*)} \nu \bar \nu$ and to $B_s \to \tau^+\tau^-$ \cite{He:2006bk}:
\begin{align}
{\cal H}_T&=-\frac{G_F}{\sqrt{2}}2s^2_W\frac{M_Z^2}{M_{Z^\prime}^2}\cot^2\theta_R V^{d*}_{Rbs}V^d_{Rbb}\bar{s}_R\gamma_\mu b_R\ \left( \bar\nu_{R3}\gamma^\mu  \nu_{R3} -\bar\tau_R \gamma^\mu \tau_R\right)\nonumber \\
{\cal H}_L&=-\frac{G_F}{\sqrt{2}}\frac{\alpha}{\pi}\frac{M_Z^2}{M_{Z^\prime}^2}\cot^2\theta_R V^{*}_{ts}V_{tb} I(\lambda_t,\lambda_H)\bar{s}_L\gamma_\mu b_L\ \left( \bar\nu_{R3}\gamma^\mu  \nu_{R3} -\bar\tau_R \gamma^\mu \tau_R\right)
\label{hfzp}
\end{align}
In the notation of \cref{effHb}, the Wilson coefficients that result in this model are thus: 
\begin{align}
C_{L}^{\prime~\tau\tau} =-C_9^{\tau\tau}=-C_{10}^{\tau\tau}&= 2\left(\frac{m_Z^2}{m_{Z^\prime}^2}\right)\cot^2\theta_R I(\lambda_t,\lambda_H)\nonumber \\
C_{R}^{\prime~\tau\tau}=-C_{9^\prime}^{\tau\tau}=-C_{10^\prime}^{\tau\tau} &= 4\left(\frac{V^{d*}_{Rbs} V^d_{Rbb}}{V_{tb}V^\star_{ts}}\right)\left(\frac{m_Z^2}{m_{Z^\prime}^2}\right)\cot^2\theta_R\frac{\pi s^2_W}{\alpha}.
\label{zpres}
\end{align}
The first operator in \cref{hfzp} originates in a flavour changing tree-level exchange of the $Z^\prime$. The parameters that appear in this result are: the $Z^\prime$ mass; a ratio parameterising the strength of the new interaction relative to the weak interaction, $\cot\theta_R$; and two elements of the matrix that rotates the down-type quarks between the weak and the mass bases. The second operator in \cref{hfzp} arises from a new penguin diagram and depends on details of the scalar sector through the Inami-Lim function $I(\lambda_t,\lambda_H)$ \cite{He:2004it}. The existing constraints on these parameters can be summarised as:
\begin{itemize}
\item A combination of perturbative unitarity \cite{He:2002ha} and LHC non-production of $Z^\prime$ from $b\bar{b}$ annihilation \cite{Hayreter:2019dzc} restrict the overall strength of the new interaction to
\begin{align}
\left(\frac{m_Z^2}{m_{Z^\prime}^2}\right)\cot^2\theta_R\lesssim 0.15.
\label{lhczp}
\end{align}
\item $B_s$ mixing constrains $|V^d_{Rbs} V^{d*}_{Rbb} |$ and $I(\lambda_t,\lambda_H)$ \cite{He:2006bk}. Both the SM calculation and the experimental situation regarding $\Delta M_{B_s}$ have  changed significantly so we repeat that analysis here.
\end{itemize}
In terms of the parameters of interest, the effective Hamiltonian below the $Z^\prime$ scale is:
\begin{align}
{\cal H}=& \frac{G_F}{\sqrt{2}}\left(\frac{m_Z^2}{m_{Z^\prime}^2}\right)\cot^2\theta_R\left(
\left(\frac{\alpha}{2\pi s_W}V^*_{tb}V_{ts}\right)^2I(\lambda_t,\lambda_H)^2 ~{\cal O}_{LL}\right.\nonumber \\
&+\left.
s_W^2 (V^d_{Rbs} V^{d*}_{Rbb})^2~{\cal O}_{RR}+2s_W\left(\frac{\alpha}{2\pi s_W}V^*_{tb}V_{ts}\right)I(\lambda_t,\lambda_H)(V^d_{Rbs} V^{d*}_{Rbb})~{\cal O}_{LR}\right) 
\label{bsmixH}
\end{align}
with the usual $\Delta B=2$ operators:
\begin{align}
{\cal O}_{LL,RR}=(\bar{s}_{L,R}\gamma_\mu b_{L,R})(\bar{s}_{L,R}\gamma_\mu b_{L,R}),\quad
{\cal O}_{LR}=(\bar{s}_{L}\gamma_\mu b_{L})(\bar{s}_{R}\gamma_\mu b_{R}).
\label{mixops}
\end{align}
QCD renormalisation group running modifies the Wilson coefficients and introduces one more operator at the $b$ scale, ${\cal O}_{SLR}=(\bar{s}_{R} b_{L})(\bar{s}_{L} b_{R})$. Making use of {\tt flavio} once more, we find
\begin{equation}
\frac{\Delta M_{B_s}}{(\Delta M_{B_s})_{SM}}\approx 1+1.5\times 10^{-4}I(\lambda_t,\lambda_H)^2-8.2~I(\lambda_t,\lambda_H) |V^d_{Rbs} V^{d*}_{Rbb} |+3282.~|V^d_{Rbs} V^{d*}_{Rbb} |^2.
\label{bsfit}
\end{equation}
The current experimental average $\Delta M_{B_s}=(17.741\pm 0.020){\rm~ps}^{-1}$ \cite{ParticleDataGroup:2020ssz} combined with a recent SM prediction $(\Delta M_{B_s})_{SM}=(18.4^{+0.7}_{-1.2}){\rm~ps}^{-1}$ \cite{DiLuzio:2019jyq} results in $\Delta M_{B_s}/(\Delta M_{B_s})_{SM}=0.96^{+0.4}_{-0.6}$ and we compare this ratio to the prediction of \cref{bsfit} in \cref{f:bsmix} for $I(\lambda_t,\lambda_H)=0$ and $I(\lambda_t,\lambda_H)=3$. These two values were chosen because a scan over the parameters in the model \cite{He:2004it} suggests $-2\lesssim I(\lambda_t,\lambda_H)\lesssim 3$ as a range for the Inami-Lim function. The allowed range for $C_{L,R}^{\prime\tau\tau}$, assuming that $V^d_{Rbs} V^{d*}_{Rbb}$ is real, is then showed in the right panel of \cref{f:bsmix}. The key point is that the tree level $Z^\prime$ exchange tends to increase $\Delta M_{B_s}$ over its SM value and this is severely constrained by current data. It is the new penguin contribution to ${\cal O}_{LR}$ and ${\cal O}_{SLR}$ that allows $\Delta M_{B_s}$ to drift below its SM value. As can be seen from \cref{bsfit}, allowing $V^d_{Rbs} V^{d*}_{Rbb}$ to have a phase can augment the allowed parameter range but a complete phenomenological study of this general case is beyond the scope of the present work.

\begin{figure}[!htb]
\begin{center}
\centering{\includegraphics[width=7cm]{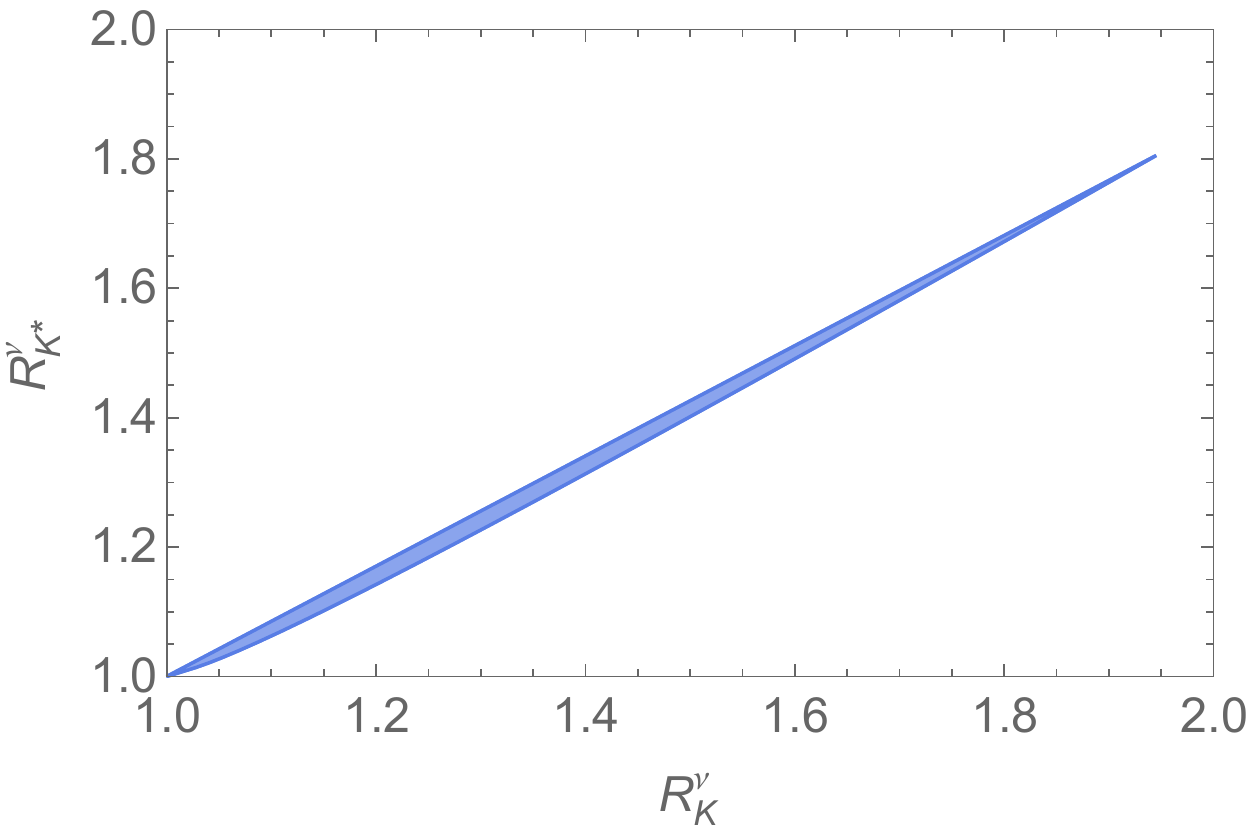}\includegraphics[width=7cm]{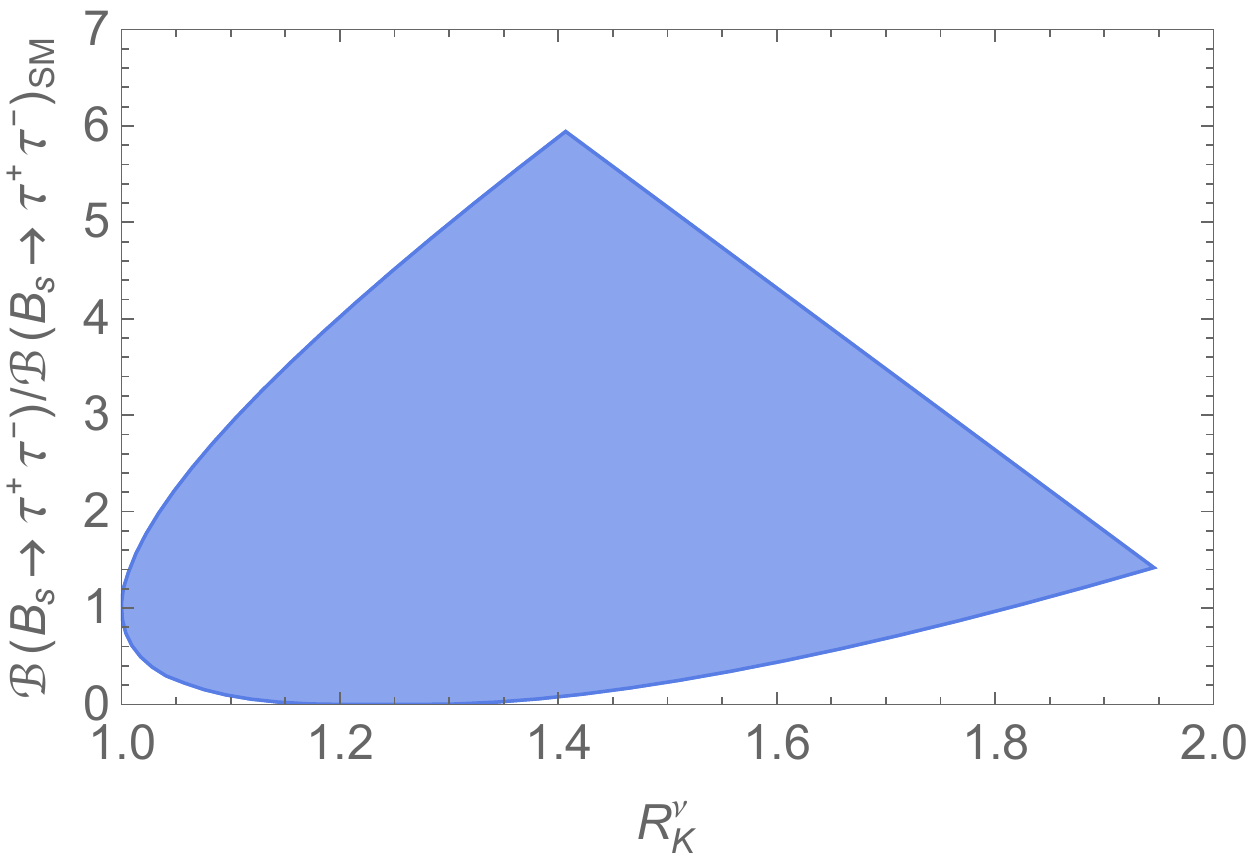}}
\end{center}
\caption{Correlation between $R^\nu_K$ and $R^\nu_{K^*}$ (left panel) and between $\mathcal{B}(B_s\to \tau^+\tau^-)$ and $R^\nu_K$ (right panel) for the  parameter space allowed by $B_s$ mixing as shown in \cref{f:bsmix}.}
\label{f:bstt}
\end{figure}

The allowed region in $C_{L,R}^{\prime\tau\tau}$ shown in \cref{f:bsmix} results in an increase of the $B\to K^{(*)}\nu\bar\nu$ rates over their SM value and this is shown in the left panel of \cref{f:bstt}. The figure indicates a near perfect correlation between the two neutrino modes and the largest values, near two, are obtained for the upper-right corner of the region in \cref{f:bsmix}. 
Interestingly, the same model with the constraint of \cref{lhczp} can no longer enhance $K\to \pi \nu \bar\nu$ modes by more than a few percent.\footnote{In the notation of \cite{He:2018uey} it predicts $\tilde{X}\leq 0.1$.} The main difference between these two cases is the strong constraint on $V^{d*}_{Rbd}$ (from $B_d$ mixing) that enters $K\to \pi \nu \bar\nu$ in place of $V^{d*}_{Rbb}$ which can be close to one. 

The model predicts through \cref{zpres} a correlation between $B\to K^{(*)}\nu\bar\nu$ modes and $B_s \to \tau^+  \tau^-$. The latter currently only has a weak experimental limit from LHCb $\mathcal{B}(B_s\to \tau^+\tau^-)\leq 6.8\times 10^{-3}$ at 95\%c.l.  \cite{Aaij:2017xqt}. The prediction can be written in a very simple form  because only the hadronic axial vector current contributes,
\begin{align}
\frac{\mathcal{B}(B_s\to \tau^+\tau^-)}{\mathcal{B}(B_s\to \tau^+\tau^-)_{SM}}= \left(1+\frac{(C_{10}^{\tau\tau}-C_{10^\prime}^{\tau\tau})}{C_{10}^{\tau\tau SM}}\right)^2.
\end{align}
The allowed parameter region seen in \cref{f:bsmix}, combined with $C_{10}^{\tau\tau SM}=-Y(x_t)/s_W^2\sim -4.3$ \cite{Buchalla:1998ba}, implies that the model allows  
\begin{equation}
\frac{\mathcal{B}(B_s\to \tau^+\tau^-)}{\mathcal{B}(B_s\to \tau^+\tau^-)_{SM}}\lesssim 6.
\end{equation}
This can be read off \cref{f:bstt} which illustrates the correlation with  $R^\nu_K$. On the other hand, the model also allows for  $(C_{10^\prime}^{\tau\tau}-C_{10}^{\tau\tau})\sim -4$ where the $\mathcal{B}(B_s\to \tau^+\tau^-)$ is significantly suppressed with respect to its SM value.

The corresponding Wilson coefficients affecting the modes $b\to s \mu^+\mu^-$ exhibiting anomalies, $C_{9,10,9^\prime,10^\prime}^{\mu\mu}$ are suppressed with respect to$C_{9,10,9^\prime,10^\prime}^{\tau\tau}$ by factors $|V^\ell_{R3\mu}|^2$ which can be very small \cite{He:2012zp}. For this reason, this model yields predictions for $b\to s \mu^+\mu^-$ processes that are very similar to the SM. Correlations between these dimuon modes and $b\to s \nu\bar\nu$ modes have also been explored in other models \cite{Altmannshofer:2009ma,Descotes-Genon:2020buf}.

\section{Models with leptoquarks}

In this section we consider models that can increase $R^\nu_{K^{(*)}}$ by producing final states with neutrino pairs of different lepton flavour, but with only the three SM neutrinos. The starting point is then scalar $S$ and vector $V$ leptoquarks with  couplings to SM fermions which include a left-handed neutrino $\nu_L$ of any flavour. They are~\cite{Davies:1990sc,Davidson:1993qk} ,
\begin{align}
{\cal L}_S =& \lambda_{LS_0} \bar q^c_L i \tau_2 \ell_L S_0^\dagger + \lambda_{L\tilde S_{1/2}} \bar d_R \ell_L \tilde S^\dagger_{1/2} + \lambda_{LS_1}\bar q^c_L i\tau_2 \vec \tau \cdot \vec S^\dagger_1 \ell_L + {\rm ~h.~c}.\;,\nonumber\\
{\cal L}_V =&  \lambda_{L V_{1/2}} \bar d_R^c \gamma_\mu \ell_L  V^{\dagger \mu}_{1/2} + \lambda_{LV_1}\bar q_L \gamma_\mu \vec \tau\cdot  \vec V^{\dagger \mu}_1 \ell_L + {\rm ~h.~c.}\;,
\end{align}
where the leptoquark fields and their transformation properties under the SM group are given by
\begin{align}
S^\dagger_0 &= S_0^{1/3} : (\bar 3, 1, 1/3) \nonumber \\
\tilde S_{1/2}^\dagger &= \left ( \tilde S_{1/2}^{-1/3}, \tilde S_{1/2}^{2/3} \right ): ( 3, 2,1/6) \nonumber\\
\vec \tau\cdot \vec S_1^{\dagger}  &=
\left (
\begin{array}{cc} 
S^{1/3}_1&\sqrt{2} S^{4/3}_1\\ 
\sqrt{2} S^{-2/3}_1&-S^{1/3}_1
\end{array}
\right ): (\bar 3, 3, 1/3)\nonumber\\
V_{1/2}^\dagger  &=\left ( V_{1/2}^{1/3}, V_{1/2}^{4/3} \right ): ( \bar 3, 2, 5/6) \nonumber\\
\vec \tau\cdot \vec V_1^{\dagger}  &=
\left (
\begin{array}{cc} 
V^{2/3}_1&\sqrt{2} V^{5/3}_1\\ 
\sqrt{2} V^{-1/3}_1&-V^{2/3}_1
\end{array}
\right ): ( 3, 3, 2/3)
\end{align}

Exchange of these particles at tree-level, assuming leptoquark multiplets that are degenerate in mass,  generates the following effective Lagrangian 
\begin{align}
{\cal L}_{eff} &= \frac{\lambda^{ij}_{LS_0} \lambda^{* kl}_{LS_0}}{ 2 m_{S_0}^2}  
\left (
\bar d_{L k}\gamma_\mu d_{L i} \bar \nu_{L_l} \gamma^\mu \nu_{Lj}  + \bar u_{L k}\gamma_\mu u_{L i} \bar e_{L_l} \gamma^\mu e_{Lj} 
- \bar u_{L k}\gamma_\mu d_{L i} \bar e_{L_l} \gamma^\mu \nu_{Lj} - \bar d_{L k}\gamma_\mu u_{L i} \bar \nu_{L_l} \gamma^\mu e_{Lj} 
\right ) \nonumber\\
&+ \frac{\lambda^{ij}_{LS_1} \lambda^{* kl}_{LS_1}}{ 2 m_{S_1}^2}   
\left ( \bar d_{L k}\gamma_\mu d_{L i} \bar \nu_{L_l} \gamma^\mu \nu_{Lj}  + 2 \bar d_{L k}\gamma_\mu d_{L i} \bar e_{L_l} \gamma^\mu e_{Lj}  + 2 \bar u_{L k}\gamma_\mu u_{L i} \bar \nu_{L_l} \gamma^\mu \nu_{Lj} +\bar u_{L k}\gamma_\mu u_{L i} \bar e_{L_l} \gamma^\mu e_{Lj} \right )
\nonumber \\
&- \frac{\lambda^{kj}_{L V_1} \lambda^{* il}_{LV_1}}{ m_{V_1}^2} 
\left ( 2 \bar d_{L k}\gamma_\mu d_{L i } \bar \nu_{L_l} \gamma^\mu \nu_{Lj}  + \bar d_{L k}\gamma_\mu d_{L i} \bar e_{L_l} \gamma^\mu e_{Lj}  +  \bar u_{L k}\gamma_\mu u_{L i } \bar \nu_{L_l} \gamma^\mu \nu_{Lj}  + 2 \bar u_{L k}\gamma_\mu u_{L i} \bar e_{L_l} \gamma^\mu e_{Lj}
\right ) \nonumber \\
&-\frac{\lambda^{kj}_{L\tilde S_{1/2}} \lambda^{* il}_{L\tilde S_{1/2}}}{ 2 m_{S_{1/2}}^2}
\left ( \bar d_{R k}\gamma_\mu d_{R i} \bar \nu_{L_l} \gamma^\mu \nu_{Lj}  + \bar d_{R k}\gamma_\mu d_{R i } \bar e_{L_l} \gamma^\mu e_{Lj} 
\right )
\nonumber \\
&+\frac{\lambda^{ij}_{L V_{1/2}} \lambda^{* kl}_{L V_{1/2}}}{ m_{V_{1/2}}^2}
\left ( \bar d_{R k}\gamma_\mu d_{R i} \bar \nu_{L_l} \gamma^\mu \nu_{Lj} + \bar d_{R k}\gamma_\mu d_{R  i} \bar e_{L_l} \gamma^\mu e_{Lj} \right ). 
\label{lqex}
\end{align}
If the fermions in \cref{lqex} are in their weak eigenstate basis, rotation to the mass eigenstate basis will introduce mixing angles. Here we will work with $\lambda^{ij}_I$ defined in a basis in which the down-type fermions are already mass eigenstates \cite{Deshpande:2012rr}. The $\nu$ and up-type quarks need to be further rotated by 
$u_{Lk} = (V^*_{KM})_{km} u_{Lm}$, and $\nu_{Lk} = (V_{PMNS})_{km} \nu_{Lm}$ respectively. However, since the neutrino flavour is not measured, working in either their weak or mass basis yields the same results. Collecting the Wilson coefficients for \cref{eq:ops} gives,
\begin{align}
C_L^{ij}&=\frac{\pi}{ \sqrt{2} \alpha G_F V_{tb}V_{ts}^*} \left (\frac{\lambda^{3j}_{LS_0} \lambda^{* 2i}_{LS_0}}{ 2 m_{S_0}^2} 
+ \frac{\lambda^{3j}_{LS_1} \lambda^{* 2i}_{LS_1}}{ 2 m_{S_1}^2} - 2  \frac{\lambda^{2j}_{L V_1} \lambda^{* 3i}_{LV_1}}{ m_{V_1}^2} 
\right)
,\nonumber\\
C_R^{ ij}& =C_{9^\prime}^{ij} = - C_{10^\prime }^{ij} = \frac{\pi}{ \sqrt{2} \alpha G_F V_{tb}V_{ts}^*}\left (- \frac{\lambda^{2j}_{L\tilde S_{1/2}} \lambda^{* 3i}_{L\tilde S_{1/2}}}{ 2 m_{S_{1/2}}^2} + \frac{\lambda^{3j}_{L V_{1/2}} \lambda^{* 2i}_{L V_{1/2}}}{ m_{V_{1/2}}^2}
\right),
\nonumber\\
C_9^{ij} &=- C_{10}^{ij} =\frac{\pi}{ \sqrt{2} \alpha G_F V_{tb}V_{ts}^*}\left (  \frac{\lambda^{3j}_{LS_1} \lambda^{* 2i}_{LS_1}}{ m_{S_1}^2} -   \frac{\lambda^{2j}_{L V_1} \lambda^{* 3i}_{LV_1}}{ m_{V_1}^2}
\right).
\label{lqres}
\end{align}

All of these leptoquarks contribute to $R^\nu_{K^{(*)}}$ but their contributions are correlated with different modes \cite{He:2019xxp,He:2019iqf,Su:2019tjn,Mandal:2019gff}. We begin with the lepton flavour number violating case which adds incoherently to the SM values for $R^\nu_{K^{(*)}}$. There are several  CLFV modes with existing experimental upper bounds and we list them in \cref{tab:explfc}. The corresponding predictions using \cref{lqres} are
\begin{align}
\mathcal{B}(B_s\to   e^\pm\mu^\mp) &\approx .98 \left(C_{10}^{\mu e~2}+C_{10^\prime}^{e\mu~2}+C_{10^\prime}^{\mu e~2}+C_{10}^{e\mu~2}\right)\times 10^{-10} \nonumber\\
\mathcal{B}(B_s\to \mu^\pm \tau^\mp) &\approx .22 \left(C_{10}^{\mu\tau~2}+C_{10^\prime}^{\tau\mu~2}+C_{10^\prime}^{\mu\tau~2}+C_{10}^{\tau\mu~2}\right)\times 10^{-7} \nonumber\\
\mathcal{B}(B^+\to K^+ e^-\mu^+) &\approx 0.18\left( (C_{10}^{\mu e}+C_{10^\prime}^{\mu e})^2+(C_{9}^{\mu e}+C_{9^\prime}^{\mu e})^2\right)\times 10^{-7}\nonumber \\
\mathcal{B}(B^+\to K^+\tau^+ e^-) &\approx 0.114\left( (C_{10}^{e\tau}+C_{10^\prime}^{e\tau})^2+(C_{9}^{e\tau}+C_{9^\prime}^{e\tau})^2\right)\times 10^{-7} \nonumber\\
\mathcal{B}(B^+\to K^+\tau^+ \mu^-) &\approx \left(0.117 (C_{10}^{\mu\tau}+C_{10^\prime}^{\mu\tau})^2+0.111 (C_{9}^{\mu\tau}+C_{9^\prime}^{\mu\tau})^2\right)\times 10^{-7} \nonumber\\
\mathcal{B}(B^+\to K^{*+ }e^-\mu^+) &\approx 0.18\left( (C_{10}^{\mu e}+C_{10^\prime}^{\mu e})^2+(C_{9}^{\mu e}+C_{9^\prime}^{\mu e})^2\right)\times 10^{-7}\nonumber \\
\mathcal{B}(B^0\to K^{*0} e^-\mu^+) &\approx 0.18\left( (C_{10}^{\mu e}+C_{10^\prime}^{\mu e})^2+(C_{9}^{\mu e}+C_{9^\prime}^{\mu e})^2\right)\times 10^{-7}
\label{fvbs}
\end{align}
The best current experimental bounds on these modes as given in \cite{ParticleDataGroup:2020ssz} are listed in \cref{tab:explfc} along with the constraints they impose on the Wilson coefficients taken one non-zero at a time.
\begin{table}[h]
	\centering
	\begin{tabular}{|l|l|c|c|}\toprule
		\textbf{Mode} & \textbf{90\% c.l} &  one $|C_i^{\ell\ell^\prime}|\neq 0$ &$C_{9^\prime}^{\ell\ell^\prime} = - C_{10^\prime }^{\ell\ell^\prime}$ or \\
		& & at a time &   $C_{9}^{\ell\ell^\prime} = - C_{10 }^{\ell\ell^\prime}$ \\
		 \midrule
		$\mathcal{B}(B_s\to e^\pm\mu^\mp)$ &  $5.4\times 10^{-9}$& 7.4 & 7.4\\
		$\mathcal{B}(B_s\to \mu^\pm\tau^\mp)$ & $4.2\times 10^{-5}$& 44 & 44\\
		$\mathcal{B}(B^+\to K^+ e^-\mu^+)$ & $6.4\times 10^{-9}$& 0.6 & 0.4 \\
		$\mathcal{B}(B^+\to K^+ e^-\tau^+)$ & $1.5\times 10^{-5}$ & 36 & 25\\
		$\mathcal{B}(B^+\to K^+ \mu^-\tau^+)$ & $2.8\times 10^{-5}$   & 49 & 35\\
		$\mathcal{B}(B^+\to K^{* +}e^-\mu^+)$ & $9.9\times 10^{-7}$   & 7.4   &5.2 \\
		$\mathcal{B}(B^0\to K^{* 0}e^-\mu^+)$ & $1.2\times 10^{-7}$   &  2.6  & 1.8\\
		 \bottomrule
	\end{tabular}
		\caption{Current experimental upper bounds on lepton flavour changing modes and the limits they imply for the coefficients $C_i^{\ell\ell^\prime}$ of \cref{fvbs} taken one non-zero at a time for the corresponding lepton flavour indices. The last column shows the upper bound on $|C_i^{\ell\ell^\prime}|$ assuming that  $C_{9^\prime}^{ij} = - C_{10^\prime }^{ij}\neq 0$, $C_{9}^{ij} = - C_{10 }^{ij}=0$ or $C_{9^\prime}^{ij} = - C_{10^\prime }^{ij}=0$, $C_{9}^{ij} = - C_{10 }^{ij}\neq 0$ as per \cref{lqres}.}
	\label{tab:explfc}
\end{table}
The minimal set of Wilson coefficients consistent with the leptoquark origin of \cref{lqres} implies more than one non-zero Wilson coefficient at a time, either $C_{9^\prime}^{ij} = - C_{10^\prime }^{ij}\neq 0$, $C_{9}^{ij} = - C_{10 }^{ij}=0$ or $C_{9^\prime}^{ij} = - C_{10^\prime }^{ij}=0$, $C_{9}^{ij} = - C_{10 }^{ij}\neq 0$. Both situations result in the same bound due to the symmetry between primed and unprimed coefficients in \cref{fvbs}. Without additional assumptions  on the leptoquark couplings, in particular allowing $C_i^{\ell\ell^\prime}$  to differ from $C_i^{\ell^\prime\ell}$  the tightest bounds that follow in this case are shown in the last column of \cref{tab:explfc}.

To explore the connection with $R^\nu_{K^{(*)}}$ it is useful to consider \cref{lqres} for each leptoquark multiplet separately. We see that $S_0$ only produces $C_L^{ij}$ and is therefore not correlated with CLFV modes. $S_1$ and $V_1$ generate $C_L^{ij}$ and $C_{9,10}^{ij}$ whereas $S_{1/2}$ and $V_{1/2}$ induce  $C_R^{ij}$ and $C_{9^\prime,10^\prime}^{ij}$ resulting in all cases in $R_{K}^\nu=R_{K^{*}}^\nu$. To study the numerical implications of these predictions we consider one lepton flavour pair at a time and present the results in \cref{t:lqresults}. These numbers indicate that the CLFV are currently less restrictive on these leptoquark couplings than  $R_{K^{(*)}}^\nu$, \cref{Rratios}, except for the $e \mu$ modes.
\begin{table}[h]
\centering
\begin{tabular}{|l|c|c|c|c|c|c|c}\toprule
\textbf{LQ} & \multicolumn{3}{|c|}{\textbf{upper bound on $C_{L,R}^{ij}$}} & \multicolumn{3}{c|}{ \textbf{$R^\nu_K=R^\nu_{K^*}$} } \\ 
&\multicolumn{3}{|c|}{} & $\mu e$&$e\tau$&$\mu\tau$ \\
\midrule
$S^0$ & $|C_L^{\mu e}|\lesssim 0.4$ & $|C_L^{e \tau}|\lesssim 26$ & $|C_L^{\mu \tau}|\lesssim 35$ & 1.001&6.4&11 \\
$\tilde S_{1/2}$ & $|C_R^{\mu e}|\lesssim 0.4$ & $|C_R^{e \tau}|\lesssim 26$ & $ |C_R^{\mu \tau}|\lesssim 35$ & 1.001&6.4&11 \\
${S}_1$ & $ |C_L^{\mu e}|\lesssim 0.2$ & $ |C_L^{e \tau}|\lesssim 13$ & $ |C_L^{\mu \tau}|\lesssim 18$ & 1.0003&2.4&3.5\\
$ V_{1/2}$ & $|C_R^{\mu e}|\lesssim 0.4$ & $ |C_R^{e \tau}|\lesssim 26$ & $ |C_R^{\mu \tau}|\lesssim 35$ & 1.001&6.4&11 \\
$V_1^\dagger$ & $ |C_L^{\mu e}|\lesssim 0.8$ & $ |C_L^{e \tau}|\lesssim 52$ & $|C_L^{\mu \tau}|\lesssim 70$ & 1.005&23&40 \\
 \bottomrule
	\end{tabular}
		\caption{Implications of limits on charged lepton flavour violation for $R_{K^{(*)}}^\nu$ for different leptoquarks shown separately for each lepton flavour combination.}
	\label{t:lqresults}
\end{table}

\subsection{The B anomalies}

Leptoquark models have been studied extensively in the context of the B anomalies and we comment on that here. The neutral B anomalies are observed in $b \to s \mu \bar \mu$ modes, and \cref{lqex} shows that, with the exception of $S_0$, these leptoquarks correlate $\bar{s} b\bar\nu^\mu\nu^\mu$ with $\bar{s} b\bar \mu \mu$ operators. In particular
\begin{align}
S_1 & \implies C_9^{\mu\mu}=-C_{10}^{\mu\mu}= 2 C_L^{\mu\mu} \nonumber \\
V_1 & \implies C_9^{\mu\mu}=-C_{10}^{\mu\mu}= \frac{1}{2} C_L^{\mu\mu} \nonumber \\
S_{1/2} & \implies C_{9^\prime}^{\mu\mu}=-C_{10^\prime}^{\mu\mu}= 2C_R^{\mu\mu} \nonumber \\
V_{1/2} & \implies C_{9^\prime}^{\mu\mu}=-C_{10^\prime}^{\mu\mu}= 2C_R^{\mu\mu}
\end{align}
Extensive fits to data from modes induced by $b \to s \mu \bar \mu$ indicate that $C_9^{\mu\mu}=-C_{10}^{\mu\mu}=-0.46$ \cite{Alguero:2019ptt}\footnote{The precise number varies depending on the fit, a more recent one gives -0.41 instead \cite{Altmannshofer:2021qrr}.} is a possible solution whereas $C_{9^\prime}^{\mu\mu}=-C_{10^\prime}^{\mu\mu}$ is not \cite{Descotes-Genon:2015uva}. This implies that 
\begin{equation}
R^\nu_{K^{(*)}}=\left\{\begin{array}{c c} 
1.02 & {\rm ~for~}S_1\\
1.1 & {\rm ~for~}V_1 \end{array}\right.
\end{equation}
and $R^\nu_{K^{(*)}}=1$ for $S_{1/2},~V_{1/2}$.

In a similar manner $S_0$ correlates $\bar{s} b\bar\nu^\tau\nu^i$ with $\bar{c} b\bar\tau \nu^i$ and therefore relates $R^\nu_{K^{(*)}}$ to the so-called charged B-anomalies, $R(D^{(*)})$.  We use the latest experimental and theoretical averages from \cite{Amhis:2019ckw} (adding errors in quadrature and using their arithmetic average of theoretical results) in terms of the ratios
\begin{equation}
r_D=\frac{R(D)}{R(D)_{SM}} =1.14\pm 0.10,\quad r_{D^*}=\frac{R(D)}{R(D^*)_{SM}} =1.14\pm 0.06
\label{chan}
\end{equation}

Depending on the neutrino lepton flavour, the operator will interfere or not with the SM and both cases were considered in \cite{Deshpande:2016yrv,BhupalDev:2021ipu}. The results are

\begin{align}
r_{D^{(*)}} &= |\Delta^{3,2}|^2 + |\Delta^{3,2}|^2 + |1+ \Delta^{3,3}|^2\nonumber \\
 \Delta^{3,j} & = -\frac{\sqrt{2}}{ 4 G_FV_{cb}} \sum_i V_{ci}\frac{\lambda^{i3*}_{L S_0}\lambda^{3j}_{L S_0}}{ 2 m^2_{S_0}}
 \label{chargedan}
\end{align}
The correlations simplify at leading order in CKM angles, where the term with $V_{cs}$ dominates resulting in,
\begin{equation}
 C_L^{3j}=-\frac{2\pi}{\alpha}\Delta^{3,2}_{j} =\frac{\pi}{ \sqrt{2}\alpha G_FV_{cb}} \frac{\lambda^{23*}_{L S_0}\lambda^{3j}_{L S_0}}{ 2 m^2_{S_0}}
\end{equation}
Taking one non-zero parameter at a time for this case, and assuming that the leptoquark contribution results in the central value of \cref{chan} requires $C_L^{33}\sim 54$ or $C_L^{31},C_L^{32} \sim 300$, both much larger than allowed by  $R^\nu_{K^{(*)}}$ as quantified in \cref{tab:milim}. Equivalently, the most favourable scenario from  \cref{tab:milim}, $C_L^{33}\sim 25 $ would result in 
\begin{equation}
r_{D^{(*)}} \sim 1.06
\end{equation}
More complex leptoquark scenarios have been invoked in the study of the B anomalies where it is possible to avoid a conflict with $R^\nu_{K^{(*)}}$ \cite{Bauer:2015knc,Cai:2017wry}.

\section{Summary}

We have studied the modes $B\to K^{(*)} \nu \bar \nu$ in the context of non-standard neutrino interactions. We first considered a model with an additional light neutrino that couples to a non-universal $Z^\prime$ and found that it can result in $R^\nu_{K^{(*)}}$ close to  two. The same model can also enhance $B_s\to \tau^+\tau^-$ by up to a factor six over the SM within the parameter range allowed by $B_s$ mixing and non-production of the $Z^\prime$ at LHC.

Next we considered augmenting $R^\nu_{K^{(*)}}$ through neutrino lepton flavour violating modes. We parameterised this possibility through scalar and vector leptoquark exchange. This type of model correlates $R^\nu_{K^{(*)}}$ with CLFV modes $B_s \to \ell\ell^\prime$ and $B\to K^{(*)}\ell \ell^\prime$ and we found that the former is currently more restrictive than $e \tau$ and $\mu \tau$ CLFV modes. 

Finally we briefly commented on the correlation with the B anomalies. In this case we saw that global fits to $b \to s \mu\mu$ modes constrain $C_L^{\mu\mu}$ for $S_1$ and $V_1$ leptoquarks  so that, {\it by itself}, it cannot add more than 10\% to $R^\nu_{K^{(*)}}$. Similarly, current measurements of $R^\nu_{K^{(*)}}$ constrain the parameters of $S_0$ couplings so that $R(D^{(*)})$ can be at most 1.06.

\section*{Acknowledgments}

This work was supported in part by the Australian Government through the Australian Research Council.
XGH was supported in part by Key Laboratory for Particle Physics, Astrophysics and Cosmology, Ministry of Education, Shanghai Key Laboratory for Particle Physics and Cosmology (Grant No. 15DZ2272100), 
in part by the NSFC (Grant Nos. 11735010, 11975149, and 12090064), and also supported in part by the MOST (Grant No. MOST 106-2112-M-002-003-MY3 ).

\bibliography{refs.bib}

\end{document}